\pdfoutput=1
\documentclass[acmsmall,nonacm]{acmart}
\usepackage{listings}
\usepackage{enumitem}
\usepackage{hyperref}
\usepackage{url}
\usepackage[linesnumbered,lined,ruled,commentsnumbered]{algorithm2e}
\usepackage{subcaption}
\usepackage{multirow}

\renewcommand{\epsilon}{\varepsilon}
\renewcommand{\emptyset}{\varnothing}

\begin{document}

\title{Optimization of the Context-Free Language Reachability Matrix-Based Algorithm}
\author{Ilia Muravev}
\email{muravjovilya@gmail.com}
\affiliation{
  \institution{Saint Petersburg State University}
  \streetaddress{7/9 University Embankment}
  \city{St. Petersburg}
  \country{Russia}
  \postcode{199034}
}
\thanks{Research advisor: Semyon Grigorev, s.v.grigoriev@spbu.ru.}
\date{November 2023}

\setcopyright{acmcopyright}
\copyrightyear{2023}
\acmYear{2023}
\acmDOI{XXXXXXX.XXXXXXX}



\begin{abstract}
Various static analysis problems are reformulated as instances of the Context-Free Language Reachability (CFL-r) problem. One promising way to make solving CFL-r more practical for large-scale interprocedural graphs is to reduce CFL-r to linear algebra operations on sparse matrices, as they are efficiently executed on modern hardware. In this work, we present five optimizations for a matrix-based CFL-r algorithm that utilize the specific properties of both the underlying semiring and the widely-used linear algebra library SuiteSparse:GraphBlas. Our experimental results show that these optimizations result in orders of magnitude speedup, with the optimized matrix-based CFL-r algorithm consistently outperforming state-of-the-art CFL-r solvers across four considered static analyses.
\end{abstract}

\maketitle

\section{Introduction}

Context-Free Language Reachability (CFL-r) is the problem of finding pairs of vertices in an edge-labeled graph that are connected by a path, with the condition that the labels along this path spell a word from a Context-Free Language (CFL) defined by a given Context-Free Grammar (CFG). A wide range of problems is reduced to CFL-r, including, among others, alias analysis~\cite{AliasAnalysis}, points-to analysis~\cite{Gigascale}, value-flow analysis~\cite{ValueFlow}, and fixing compilation errors~\cite{FixCompilation}. Furthermore, CFL-r is also used to approximate solutions for some undecidable problems, for example, points-to analysis that is simultaneously context-, field-, and flow-sensitive~\cite{SPDS, Undecidable}.

Since Thomas Reps et al.~\cite{Reps} proposed using CFL-r for precise interprocedural dataflow analysis, numerous CFL-r algorithms have been developed~\cite{Matrix, GLL, kron, POCR}. 
One of the promising general-purpose CFL-r algorithms is a matrix-based algorithm by Rustam Azimov et al.~\cite{Matrix}. This algorithm naturally utilizes parallel hardware and oftentimes surpasses analogous in terms of performance~\cite{RustamPHD}. However, it still takes a considerable amount of time to deal with paths with deep derivation trees and large CFGs of field- and context-sensitive analyses. In this work, we propose a set of optimizations that improve performance of the matrix-based algorithm in such cases. Our evaluation demonstrates that the optimized version outperforms such tools as POCR~\cite{POCR}, Gigascale~\cite{Gigascale}, and Graspan~\cite{Graspan}.

\section{Preliminaries}

\begin{definition}[Weak Chomsky Normal Form (WCNF)] Let $Gr = (N, \Sigma, P, S)$ be a CFG, where $N$ is a set of non-terminals, $\Sigma$ is a set of terminals, $P$ is a set of productions, and $S$ is the starting non-terminal. $Gr$ is said to be in WCNF if $P \subseteq \{a \rightarrow b \mid a \in N, b \in \Sigma \cup \{\epsilon\}\} \cup \{c \rightarrow a\ b \mid a, b, c \in N\}$.
\end{definition}

\begin{definition}[Reachability Semiring of CFG]
    Let $Gr = (N, \Sigma, P, S)$ be a CFG in WCNF. In this paper, $R_{Gr} = (2^N, \cup, \otimes_{Gr}, \emptyset)$ is called a reachability semiring\footnote{$R_{Gr}$ is a semiring according to GraphBlas~\cite{Graphblas}, but not according to the common algebraic definition of a semiring.} of $Gr$, where $2^N$ is the domain, $\cup$ acts as addition, $\otimes_{Gr}$ acts as multiplication, and $A \otimes_{Gr} B = \{ c \in N \mid \exists (a, b) \in A \times B\ ((c \rightarrow a\ b) \in P)\}$.
\end{definition}

\begin{algorithm}
\caption{Original Matrix-Based CFL-r Algorithm as Introduced in~\cite{Matrix}}
\label{algo:matrix}
\KwData{CFG $Gr = (N, \Sigma, P, S)$, graph $G = (V, E)$, where $V = 1..n$ are vertices and $E$ are edges}
\KwResult{Matrix $M \in R_{Gr}^{|V| \times |V|}$, where $M_{ij} = \{ x \in N \mid \exists \text{ path from $i$ to $j$ derivable from $x$} \}$}

$Gr \gets \text{CFG in WCNF equivalent to CFG $Gr$}$\;

$M \gets$ zero $|V| \times |V|$ matrix over $R_{Gr}$\tcp*{Stored as $|N|$ Boolean matrices}
\ForEach{$(i, x, j) \in E$}{
    $M_{ij} \gets M_{ij} \cup \{A \mid (A \to x) \in P\}$\;
}

\While{$M$ is changing}{
    $M \gets M \cup (M \cdot_{R_{Gr}} M)$\tcp*{$\cdot_{R_{Gr}}$ is a matrix multiplication in semiring $R_{Gr}$}
    \label{algo:matrix:closure}
} 

\end{algorithm}

\section{Optimizations}

Following optimizations have been proposed, implemented, and evaluated.
\begin{enumerate}
    \item The bottleneck in Algorithm \ref{algo:matrix} is at $\cdot_{R_{Gr}}$ on line \ref{algo:matrix:closure}. To address this, $M \cdot_{R_{Gr}} M$ is replaced with $(M_{old} \cdot_{R_{Gr}} \Delta M) \cup (\Delta M \cdot_{R_{Gr}} M)$, where $\Delta M = M \setminus M_{old}$ (element-wise set difference) and $M_{old}$ is the value of the variable $M$ during the previous iteration of the \texttt{while} loop. Initially, $M_{old} = \emptyset^{|V| \times |V|}$. The rationale here is that $(M_{old} \cdot_{R_{Gr}} \Delta M) \cup (\Delta M \cdot_{R_{Gr}} M) \cup (M_{old} \cdot_{R_{Gr}} M_{old}) = M \cdot_{R_{Gr}} M$ and $M_{old} \cdot_{R_{Gr}} M_{old}$ has already been added to $M$ on the previous iteration.
    \item Now, during most iterations, $\Delta M$ is highly sparse, while $M_{old}$ and $M$ are less sparse. The multiplication of a highly sparse matrix with a less sparse one is considerably faster in row-major format when the sparser matrix is on the left and in column-major format when the sparser matrix is on the right~\cite{Graphblas}. Therefore, the next optimization involves maintaining two copies of $M$ in both row- and column-major formats and computing $M_{old} \cdot_{R_{Gr}} \Delta M$ and $\Delta M \cdot_{R_{Gr}} M$ using column-by-column and row-by-row (hyper)sparse matrix multiplication techniques, respectively~\cite{SpGEMM}. In practice, non-terminals that don't occur in place of $a$ in productions like $c \rightarrow a\ b$ don't need to be stored in the column-major copy of $M$, and those that don't occur in place of $b$ don't need to be stored in the row-major copy of $M$.
    \item \label{opt:iadd} The next bottleneck is the element-wise union on line \ref{algo:matrix:closure}, which becomes problematic as adding a highly sparse matrix to $M$ causes a complete reconstruction of $M$. To cope with this, $M$ is no longer stored as a matrix. Instead, we now use a set of matrices $\widetilde{M} = \{M_1, M_2, \dots, M_p\}$ such that $M = \bigcup \widetilde{M}, \forall A, B \in \widetilde{M}\ (nnz(A) < nnz(B) \implies b\ nnz(A) < nnz(B))$, where $b > 1$ is a hyperparameter and $nnz(X)$ is the number of non-zero elements in  matrix $X$.
    The element-wise union on line~\ref{algo:matrix:closure} is computed by adding $\Delta M$ to $\widetilde{M}$ as a set element and replacing $\widetilde{M}$ with $\widetilde{M} \setminus \{A, B\} \cup \{A \cup B\}$ as long as there are $A, B \in \widetilde{M}$ that violate $\widetilde{M}$'s invariants.
    \item For CFGs with a large number of productions ($|P|$), the bottleneck is once again $\cdot_{R_{Gr}}$, requiring $|P|$ Boolean matrix multiplications~\cite{VALIANT}. Fortunately, such CFGs often use \emph{``indexed''} non-terminals (see examples in appendix~\ref{appendix:data}). We exploit this by storing only two Boolean block-matrices (horizontal and vertical) per entire \emph{``indexed''} non-terminal, reducing the number of Boolean matrix multiplications needed for $\cdot_{R_{Gr}}$ to the number of productions differing not only in indices. For example, $\forall i\ (AR_i \rightarrow A\ ret_i)$ now counts as one production. 
    \item \sloppy \label{opt:cfg} Finally, profiling has revealed equivalent transformations of CFGs for field-sensitive Java points-to~\cite{Gigascale} and field-insensitive C/C++ memory-alias\footnote{Transformed CFG for C/C++ memory-alias analysis is no longer usable for C/C++ value-alias analysis.}~\cite{AliasAnalysis} analyses that improve performance (see appendix~\ref{appendix:data}).
\end{enumerate}

\section{Evaluation}

All experiments were run on a PC with an Intel Core i7-6700 CPU (3.4 GHz, 4 threads, hyper-threading is turned off), and DDR4 64GB RAM, running Ubuntu 20.04 with SuiteSparse:GraphBLAS 7.4.4 and Java HotSpot(TM) 64-Bit Server VM (build 15.0.2+7-27, mixed mode, sharing) installed.

Proposed variants of Algorithm~\ref{algo:matrix} were implemented\footnote{Implementations of matrix-based algorithm variants --- \url{https://github.com/FormalLanguageConstrainedPathQuerying/CFPQ\_PyAlgo/tree/murav/optimize-matrix} (date of
access: 06.11.2023), for optimization (\ref{opt:iadd}) hyperparameter $b = 10$ is used.} and compared with state-of-the-art CFL-r implementations: POCR\footnote{POCR --- \url{https://github.com/kisslune/POCR} (date of
access: 06.11.2023).}~\cite{POCR} and Graspan\footnote{Graspan --- \url{https://github.com/Graspan/Graspan-C} (date of
access: 06.11.2023).}~\cite{Graspan}, as well as with the field-sensitive Java points-to analyzer Gigascale\footnote{Gigascale --- \url{https://bitbucket.org/jensdietrich/gigascale-pointsto-oopsla2015/src/master/} (date of access: 06.11.2023).}~\cite{Gigascale} on graphs from the CFPQ Data\footnote{CFPQ Data --- \url{https://formallanguageconstrainedpathquerying.github.io/CFPQ_Data} (date of access: 06.11.2023).} and CPU17\footnote{CPU17 (graphs for SPEC 2017 C/C++ programs) --- \url{https://github.com/kisslune/CPU17-graphs} (date of access: 06.11.2023). Graphs from the ``simplified-interdyck'' folders are used, as done in POCR~\cite{POCR}. Our evaluation excludes the time taken to construct SVFG and PEG graphs, eliminate cycles~\cite{Cycle}, and substitute variables~\cite{Subst}, since it remains the same regardless of the CFL-r implementation.} data sets and their corresponding canonical CFGs (see appendix~\ref{appendix:data}).
Algorithm~\ref{algo:matrix} with optimizations $a, b, \dots, c$ is denoted by $ma_{a b\dots c}$.
For example, $ma_{12345}$ means algorithm~\ref{algo:matrix} with all proposed optimizations.

The following analyses are considered: Field-Sensitive Java Points-To (FSJPT)~\cite{Gigascale}, Field-Insensitive C/C++ Alias (FICA)~\cite{AliasAnalysis}, Field-Sensitive C/C++ Alias (FSCA)~\cite{AliasAnalysis}, and Context-Sensitive C/C++ Value-Flow (CSCVF)~\cite{ValueFlow}. To show the worst-case performance of proposed optimizations, for each analysis, two graphs that took $ma_{12345}$ the longest time to analyze were selected. For selected graphs, the mean run time over five runs is displayed in Table~\ref{tab:run-time}. In all cases, the unbiased standard deviation estimate is less than 10\% of the mean. OOT (Out of Time) means exceeding the 10,000-second timeout, OOM (Out of Memory) refers to running out of available memory, and a dash ``-'' indicates that a particular implementation is not applicable to a problem\footnote{Gigascale is not a general CFL-r solver and only solves FSJPT, while Graspan encodes each non-terminal using a single byte and hence can't handle arbitrarily large CFGs for field- and context-sensitive analyses.}. For selected graphs, proposed optimizations preserved the correctness of the answers and increased RAM consumption by at most 70\% compared to the baseline (see appendix~\ref{appendix:ram}). 

\begin{table}
  \centering
  \caption{All-Pairs CFL-r Run Time in Seconds}
  \begin{tabular}{|c|c|c|c|c|c|c|c|c|c|}
    \hline
    Problem               & Graph        & $ma$  & $ma_1$  & $ma_{14}$     & $ma_{1234}$    & $ma_{12345}$  & POCR  & Graspan & Gigascale \\
    \hline
    \multirow{2}{*}{FSJPT}
                          & tradebeans   &  OOT  & 5075    & 51            &   20           & \textbf{1.3}  &   395 &       - &       5.1 \\
                          & tradesoap    &  OOT  & 5284    & 52            &   20           & \textbf{1.5}  &   400 &       - &      4.5 \\
    \hline
    \multirow{2}{*}{FICA}
                          & apache       & 999   & 139     & 139           & 88             & \textbf{19}   & OOM   & 2619    & -  \\
                          & postgre      & 1365  & 186     & 186           & 109            & \textbf{30}   & OOM   & 1712    & -  \\
    \hline
    \multirow{2}{*}{FSCA}
                          & imagick      & 2279  & 372     & 214           &   \textbf{137} &  \textbf{137} &  1538 &       - &         - \\
                          & perlbench    & 4544  & 1721    & \textbf{1321} &  1675          &  1675         &   OOT &       - &         - \\
    \hline
    \multirow{2}{*}{CSCVF}
                          & povray       &  OOT  & 1057    & 21.7          &   \textbf{10}  & \textbf{10}   &  28   &       - &         - \\
                          & perlbench    &  OOT  & OOT     & 195           &   \textbf{41}  & \textbf{41}   &   OOM &       - &         - \\

    \hline
  \end{tabular}
  \label{tab:run-time}
\end{table}

\section{Conclusion and future work}

The optimized version of the matrix-based CFL-r algorithm was shown to outperform analogous  across a range of considered problems. Building on this success, future work is expected to include complexity analysis as well as generalization of proposed optimizations for other algorithms\footnote{All proposed optimizations can already be transferred to single-path semantics of CFL-r by merely changing the semiring.}.

\bibliographystyle{ACM-Reference-Format}
\bibliography{matrix_CFLr_optimization}

\appendix

\section{RAM consumption}
Table~\ref{tab:ram} displays the maximum RAM usage for all conducted experiments, using the same notations as in Table~\ref{tab:run-time}.
\label{appendix:ram}

\begin{table}
  \centering
  \caption{All-Pairs CFL-r RAM Consumption in GB}
  \begin{tabular}{|c|c|c|c|c|c|c|c|c|c|}
    \hline
    Problem               & Graph             &         $ma$ & $ma_1$ &   $ma_{14}$ & $ma_{1234}$ & $ma_{12345}$   &            POCR & Graspan & Gigascale \\
    \hline
    \multirow{2}{*}{FSJPT}
                          & tradebeans        &          OOT & 2.2    &         1.9 &    1.8      &  \textbf{0.35} &   3.2           &       - &       1.0 \\
                          & tradesoap         &          OOT & 2.2    &         1.9 &    1.8      &  \textbf{0.35} &   3.2           &       - &       1.0 \\
    \hline
    \multirow{2}{*}{FICA} 
                          & apache            &           12 & 11     &          11 &     12      &   \textbf{2.9} &   OOM          &      11 &         - \\
                          & postgre           &           11 & 11     &          11 &     11      &   \textbf{3.3} &   OOM           &      12 &         - \\
    \hline
    \multirow{2}{*}{FSCA} 
                          & imagick           &           13 & 11     &          11 &     14      &      14        &   \textbf{7.3} &       - &         - \\
                          & perlbench         &  \textbf{16} & 21     &          21 &     27      &      27        &            OOT &       - &         - \\
    \hline
    \multirow{2}{*}{CSCVF} & povray            &          OOT & 1.3    &\textbf{1.1} &    1.2      &     1.2        &            9.9 &       - &         - \\
                          & perlbench         &          OOT & OOT    &\textbf{5.8} &    6.0 &     6.0             &            OOM &       - &         - \\
    \hline
  \end{tabular}
  \label{tab:ram}
\end{table}

\section{Graphs and grammars}
\label{appendix:data}

Table~\ref{tab:graph} displays the characteristics of analyzed graphs: number of vertices, edges, and distinct labels.

\begin{table}
\centering
\caption{Characteristics Of Analyzed Graphs}
\begin{tabular}{|c|c|c|c|c|}
\hline
Problem & Graph & Vertices & Edges & Distinct Edge Labels \\
\hline
\multirow{2}{*}{FSJPT}
                       & tradebeans &  439,693 & 933,938 & 31,886 \\
                       & tradesoap & 440,680 & 936,526 & 31,980 \\
\hline
\multirow{2}{*}{FICA} 
                       & apache & 1,721,418 & 3,020,822 & 4 $(a, d, \overline{a}, \overline{d})$ \\
                       & postgre & 5,203,419 & 9,357,086 & 4 $(a, d, \overline{a}, \overline{d})$ \\
\hline
\multirow{2}{*}{FSCA}  
                       & imagick & 41,652 & 111,550 & 288 \\
                       & perlbench & 38,091 & 110,874 & 110 \\
\hline
\multirow{2}{*}{CSCVF} 
                       & povray & 346,034 & 581,210 & 11,311 \\
                       & perlbench & 605,864 & 1,114,892 & 28,881 \\
\hline
\end{tabular}
\label{tab:graph}
\end{table}

In this work, as optimization~(\ref{opt:cfg}) states, CFGs for field-sensitive Java points-to~\cite{Gigascale} and field-insensitive C/C++ memory-alias~\cite{AliasAnalysis} analyses were manually transformed to WCNF (see Figures~\ref{fig:wcnf:fsjpt} and~\ref{fig:wcnf:fica}) in such a way that results in better performance compared to automatically generated WCNF (compare $ma_{1234}$ and $ma_{12345}$ in Table $\ref{tab:run-time}$).

Table~\ref{tab:cfg-per-algo} shows, for each pair of CFL-r implementation and a problem, which CFG was used. ``Embedded'' means that the representation of CFL for a specific problem is embedded into the CFL-r implementation itself, and an asterisk ``*'' indicates that the CFG is automatically\footnote{CFG normalization implementation --- \url{https://formallanguageconstrainedpathquerying.github.io/CFPQ\_Data/reference/grammars/generated/cfpq\_data.grammars.converters.cnf.html\#cfpq\_data.grammars.converters.cnf.cnf\_from\_cfg} (date of access: 06.11.2023).} normalized, because a particular CFL-r implementation only works with CFGs in WCNF.

\begin{table}
  \centering
  \caption{Used CFGs}
  \begin{tabular}{|c|c|c|c|c|c|c|}
    \hline
    Problem             & $ma$, $ma_1$ $ma_{14}$, $ma_{1234}$ & $ma_{12345}$                & POCR                         & Graspan                    & Gigascale \\
    \hline
    FSJPT               & Figure~\ref{fig:cfg:fsjpt}*         & Figure~\ref{fig:wcnf:fsjpt} & Figure~\ref{fig:cfg:fsjpt}*  & -                          & Embedded \\
    FICA                & Figure~\ref{fig:cfg:fica}*          & Figure~\ref{fig:wcnf:fica}  & Figure~\ref{fig:cfg:fica}*   & Figure~\ref{fig:cfg:fica}* & - \\
    FSCA                & Figure~\ref{fig:wcnf:fsca}          & Figure~\ref{fig:wcnf:fsca}  & Embedded                     & -                          & - \\
    CSCVF               & Figure~\ref{fig:wcnf:cscvf}         & Figure~\ref{fig:wcnf:cscvf} & Embedded                     & -                          & - \\
    \hline
  \end{tabular}
  \label{tab:cfg-per-algo}
\end{table}

\begin{figure}
\centering
\begin{subfigure}[m]{.4\textwidth}
  \centering
  \begin{alignat*}{2}
    & PT &&\to PTH\ alloc \\
    & PTH &&\to \epsilon \mid assign\ PTH \\
    & PTH &&\to load_i\ Al\ store_i\ PTH \\
    & FT &&\to \overline{alloc}\ FTH \\
    & FTH &&\to \epsilon \mid \overline{assign}\ FTH \\
    & FTH &&\to \overline{store_i}\ Al\ \overline{load_i}\ FTH \\
    & Al &&\to PT\ FT
  \end{alignat*} 
  \caption{CFG taken from CFPQ Data}
  \label{fig:cfg:fsjpt}
\end{subfigure}
\hspace{.025\textwidth}
\begin{subfigure}[m]{.5\textwidth}
  \centering
  \begin{alignat*}{2}
    & PT &&\to alloc \mid assign\ PT \mid LPFS_i\ PT\\
    & FT &&\to \overline{alloc} \mid FT\ \overline{assign} \mid FT\ SPFL_i \\
    & LPFS_i &&\to LP_i\ FS_i \\
    & LP_i &&\to load_i\ PT \\
    & FS_i &&\to FT\ store_i \\
    & SPFL_i &&\to SP_i\ FL_i \\
    & SP_i &&\to \overline{store_i}\ PT \\
    & FL_i &&\to FT\ \overline{load_i}
  \end{alignat*} 
  \caption{CFG in WCNF, introduced as optimization (\ref{opt:cfg})}
  \label{fig:wcnf:fsjpt}
\end{subfigure}
\caption{CFG for field-sensitive Java points-to analysis}
\end{figure}

\begin{figure}
\centering
\begin{subfigure}[m]{.3\textwidth}
  \centering
  \begin{alignat*}{2}
    & M  &&\to \overline{d}\ V\ d \\
    & V  &&\to \epsilon \mid V_1\ V_2\ V_3 \\
    & V_1 &&\to \epsilon \mid V_2\ \overline{a}\ V_1 \\
    & V_2 &&\to \epsilon \mid M \\
    & V_3 &&\to \epsilon \mid a\ V_2\ V_3
  \end{alignat*} 
  \caption{CFG taken from CFPQ Data}
  \label{fig:cfg:fica}
\end{subfigure}
\hspace{.025\textwidth}
\begin{subfigure}[m]{.45\textwidth}
  \centering
  \begin{alignat*}{2}
    & M &&\to N_1\ N_3 \mid N_2\ N_3 \\
    & N_1 &&\to \overline{d} \mid N_1\ \overline{a} \mid N_2\ \overline{a}\\
    & N_2 &&\to N_1\ M\\
    & N_3 &&\to d \mid a\ N_3 \mid AM\ N_3 \\
    & AM &&\to a\ M
  \end{alignat*} 
  \caption{CFG in WCNF, introduced as optimization (\ref{opt:cfg})}
  \label{fig:wcnf:fica}
\end{subfigure}
\caption{CFG for field-insensitive C/C++ alias analysis}
\end{figure}

\begin{figure}
\centering
\begin{subfigure}[m]{.3\textwidth}
  \centering
  \begin{alignat*}{2}
    & M  &&\to \overline{d}\ V\ d \\
    & V  &&\to \overline{A}\ V\ A \mid \overline{f_i}\ V\ f_i \mid M \mid \epsilon \\
    & A  &&\to a\ M? \mid \epsilon \\
    & \overline{A}  &&\to M?\ \overline{a} \mid \epsilon
  \end{alignat*} 
  \caption{CFG}
  \label{fig:cfg:fsca}
\end{subfigure}
\hspace{.025\textwidth}
\begin{subfigure}[m]{.4\textwidth}
  \centering
  \begin{alignat*}{2}
    & M  &&\to DV\ d \\
    & DV &&\to \overline{d}\ V \\
    & V  &&\to \overline{A}\ V \mid V\ A \mid FV_i\ f_i \mid M \mid \epsilon \\
    & FV_i &&\to \overline{f}_i V \\
    & A  &&\to a\ M \mid a \mid \epsilon \\
    & \overline{A} &&\to  M\ \overline{a} \mid \overline{a} \mid \epsilon
  \end{alignat*} 
  \caption{CFG in WCNF}
  \label{fig:wcnf:fsca}
\end{subfigure}
\caption{CFG for field-sensitive C/C++ alias analysis, taken from~\cite{POCR}}
\end{figure}

\begin{figure}
\centering
\begin{subfigure}[m]{.25\textwidth}
  \centering
  \begin{alignat*}{2}
    & A  &&\to A\ A \mid a \mid \epsilon \\
    & A  &&\to call_i\ A\ ret_i
  \end{alignat*} 
  \caption{CFG}
  \label{fig:cfg:cscvf}
\end{subfigure}
\hspace{.025\textwidth}
\begin{subfigure}[m]{.3\textwidth}
  \centering
  \begin{alignat*}{2}
    & A  &&\to A\ a \mid A\ AH \mid \epsilon \\
    & AH &&\to call_i\ AR_i \\
    & AR_i  &&\to A\ ret_i
  \end{alignat*} 
  \caption{CFG in WCNF}
  \label{fig:wcnf:cscvf}
\end{subfigure}
\caption{CFG for context-sensitive C/C++ value-flow analysis, taken from~\cite{POCR}}
\end{figure}

\end{document}